\documentclass[conference]{IEEEtran}
\usepackage[utf8]{inputenc}
\usepackage[T1]{fontenc}
\usepackage{hyperref}
\usepackage{url}
\usepackage{booktabs}
\usepackage{amsmath,amssymb,amsfonts}
\usepackage{nicefrac}
\usepackage{microtype}
\usepackage{lipsum}
\usepackage{graphicx}
\usepackage{natbib}

\title{A Fuzzy Approach to Project Success:\\ Measuring What Matters}

\author{
\IEEEauthorblockN{João Granja-Correia}
\IEEEauthorblockA{Universidad de Cantabria \\ Santander, España \\ jgc672@alumnos.unican.es}
\and
\IEEEauthorblockN{Remedios Hernández-Linares}
\IEEEauthorblockA{Universidad de Extremadura \\ Mérida, España \\ remedioshl@unex.es}
\and
\IEEEauthorblockN{Luca Ferranti}
\IEEEauthorblockA{Aalto University \\ Espoo, Finland \\ luca.ferranti@aalto.fi}
\and
\IEEEauthorblockN{Arménio Rego}
\IEEEauthorblockA{Católica Porto Business School \\ Porto, Portugal \\ arego@ucp.pt}
}

\begin{document}
\maketitle

\begin{abstract}
This paper introduces a novel approach to project success evaluation by integrating fuzzy logic into an existing construct. Traditional Likert-scale measures often overlook the context-dependent and multifaceted nature of project success. The proposed hierarchical Type-1 Mamdani fuzzy system prioritizes sustained positive impact for end-users, reducing emphasis on secondary outcomes like stakeholder satisfaction and internal project success. This dynamic approach may provide a more accurate measure of project success and could be adaptable to complex evaluations. Future research will focus on empirical testing and broader applications of fuzzy logic in social science.
\end{abstract}

\section{Introduction}
The concept of project success has evolved significantly, from the traditional "iron triangle" of time, cost, and scope \cite{pollack2018} to a broader perspective that recognizes its multifaceted and context-dependent nature. Modern approaches emphasize stakeholder perspectives across organizational levels, acknowledging that success extends beyond technical deliverables to include stakeholder satisfaction and broader strategic goals \cite{castro2021}.

Ika and Pinto's 2022 editorial highlights the limitations of focusing on unidimensional project metrics and advocates for including long-term sustainability measures to achieve more comprehensive assessments \cite{ika2022}. They propose developing a typology of projects and contexts to guide the selection of success models, integrating internal and external stakeholder perceptions alongside sustainability dimensions.

Effectively assessing project success requires a phronetic understanding of the project, its context, stakeholders' needs, and potential future developments \cite{clegg2014}. This complexity renders the concept inherently fuzzy, as traditional models often fail to capture real-world nuances \cite[p.~833]{PINTO2022831}.

Fuzzy logic, originating from the work of Łukasiewicz and Zadeh, provides a flexible framework for managing uncertainties, making it suitable for evaluating project success \cite{GILES1976313}. Unlike binary logic, fuzzy logic accommodates degrees of membership, enabling the representation of imprecise and uncertain information—crucial for assessing complex phenomena like project success, which involves inherent vagueness and dynamic relationships \cite{Settimo_2010}.

In social science, constructs are abstract concepts measured through observable indicators like Likert scale items, which assess attitudes or opinions by having respondents rate their agreement with statements on a symmetric scale \cite{Clark20191412}. Traditional approaches often oversimplify complex phenomena by neglecting the subtleties and context of inter-item relationships. To address these limitations in project success research, our fuzzy approach incorporates hierarchical structures and context-sensitive logic. This provides a more accurate and nuanced assessment tool, allowing for a deeper understanding of the multifaceted nature of project success.

This paper presents two contributions: a method to adapt constructs for their hierarchical and context-dependent nature, and a reinterpretation of an established project management construct to better capture the nuanced dimensions of project success, providing a more meaningful tool for future research.

\section{Theoretical Framework}
While projects are temporary endeavors aimed at achieving specific goals \cite{beckerExploratoryStudySyntactic2024}, evaluating their success involves navigating complex stakeholder views and numerous influencing factors \cite{sunBalancingProjectManagement2024}. Traditional evaluation methods often struggle to encapsulate this complexity \cite{Joslin2016613}, leading to incomplete assessments. To address these challenges, fuzzy logic, particularly the Mamdani fuzzy inference system developed by Ebrahim Mamdani in 1977, emerges as a valuable tool. By representing partial truths and accommodating diverse perspectives, this system effectively models the relationships between various project dimensions and overall success \cite{daniello2023fuzzy}, thereby potentially offering a more comprehensive evaluation framework.

\subsection{Project Success Measurement Tool}
The instrument used in this study to evaluate project success is a multidimensional construct developed by Aga et al. (2016) \cite{aga2016}. It systematically assesses project managers' perceptions across criteria such as time management, cost control, and client satisfaction. This construct is distilled into 14 items [p.~816]\cite{aga2016}, each evaluated using a five-point Likert scale. This approach allows respondents to indicate their level of agreement or satisfaction, providing a structured framework for capturing the nuanced aspects of project success.

\subsection{Five-Point Likert Scale}
Likert scales are essential in management research for assessing subjective data like attitudes and perceptions \cite{2021.637547}. However, traditional Likert scales struggle with ambiguity and assume equal intervals between points. To overcome these limitations, researchers have integrated fuzzy set theory, enhancing assessment flexibility \cite{Lalla2005577}. In this study, Likert scales are used as key input variables, quantifying subjective perspectives based on a validated construct \cite{aga2016}. We employ a five-point Likert scale to capture responses, facilitating fuzzification—converting degrees of agreement into numerical membership degrees. This ensures clarity during defuzzification, where values are transformed back into the five-point scale format. The method is adaptable to a seven-point scale with minor adjustments.

\section{Methodology}
This study employs a multi-level Mamdani fuzzy logic approach \cite{1638460} to evaluate project success, effectively capturing its nuances and subjectivity. For transparency and reproducibility, the complete implementation, including scripts, examples, documentation, and visualizations, is available on GitHub.\footnote{\url{https://github.com/joaojcorreia/FuzzyLogic_ProjectSuccess}}

\subsection{From Individual Items to a Collective Measure of Success: Fuzzy Reasoning and Defuzzification}
We retain the five-point Likert scale for both intermediate and final success measures, ensuring consistency for quantitative studies and maintaining interpretability within the original qualitative context. The process involves assessing and preserving the Likert scale for measuring success.

\begin{enumerate}
\item\textbf{Grouping and Fuzzification}: Items are grouped by dimensions and fuzzified based on predefined principles, converting discrete data into fuzzy values to represent relationships more nuancedly \cite{Karnik1999643}.
\item \textbf{Defuzzification of Aggregated Dimensions}: After fuzzification, dimensions are defuzzified to produce crisp intermediate success measures, ensuring clarity and actionability.
\item \textbf{Fuzzification and Defuzzification for Final Success Score}: Aggregated dimensions are further fuzzified to account for interrelationships, then defuzzified to generate a final success score reflecting the collective impact on project success.
\end{enumerate}

\subsection{Decomposing the Construct: Hierarchical Structure}
The initial step involves decomposing the overall project success construct, the 14 project success items into three key dimensions:

\begin{enumerate}
\item \textbf{Project Management Success}: This group focuses on the classic iron triangle (time and budget constraints) and stakeholder satisfaction during implementation, assessing internal project execution and management practices.
\item \textbf{Measure of Project Impact Success}: This dimension evaluates the project's impact and benefits for users and beneficiaries, emphasizing problem-solving, performance improvement, and lasting positive effects.
\item \textbf{Stakeholder Satisfaction}: This dimension assesses satisfaction levels of key stakeholders, including end-users, beneficiaries, and donors, evaluating the project's success in meeting expectations and fostering positive perceptions.
\end{enumerate}

\subsection{Framework for Integrating Project Success Metrics}
The system assesses project success in two levels. The first level aggregates individual success items into three dimensions: Project Management Success, Project Impact Success, and Stakeholder Satisfaction, each evaluated through a separate Mamdani fuzzy inference system. Success items are mapped to fuzzy sets (failure, neutral, success) and combined via expert-based rules to produce an aggregated success level for each dimension. The second level combines these dimension scores using another Mamdani inference system, accounting for their relative importance and interdependencies, to generate a single crisp value representing overall project success.

\begin{figure}[h]
  \centering
  \includegraphics[width=0.5\textwidth]{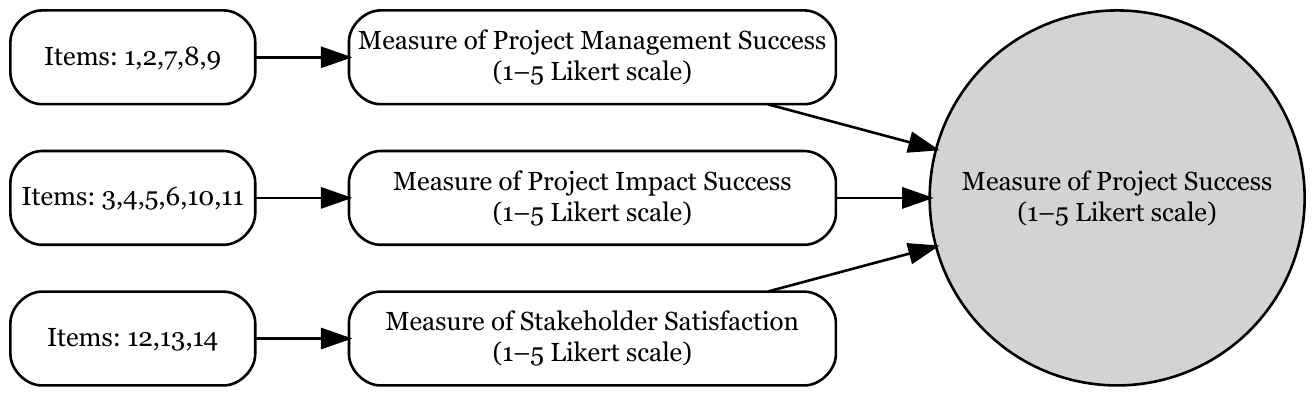}
  \caption{Fuzzification and Defuzzification Process}
  \label{fig:fig1}
\end{figure}

\section{Conclusion}
This study introduces a novel fuzzy logic approach to capture the hierarchical and context-dependent nature of project success. It adapts a validated construct widely used in project management research, offering a reinterpreted project success construct aimed at delivering more meaningful research outputs.

While this model provides valuable insights, it requires empirical validation and exploration of alternative aggregation structures for Project Management Success, Project Impact Success, and Stakeholder Satisfaction. Future research should replicate existing studies using the same construct to compare results with established methods. This will help validate the fuzzy measure, reveal stronger causal relationships, and enhance model robustness and applicability. Validation with relevant project stakeholders will also be essential to refine the construct and capture evolving success dimensions.

\bibliographystyle{unsrt} 
\bibliography{references} 

\end{document}